\documentclass{elsart}
 \usepackage{graphicx}
 \usepackage{amssymb,amsmath}
\begin{document}
\begin{frontmatter}
\title{Collisional Effects on HFS Transitions of Antiprotonic Helium}
\author{G.Ya.Korenman\corauthref{cor}},
 \corauth[cor]{Corresponding author}
\ead{korenman@anna19.npi.msu.su}
\author{N.P.Yudin},
\and
\ead{yudin@helene.sinp.msu.ru}
\author{S.N.Yudin}
\ead{yudins@anna19.npi.msu.su}
\address{Institute of Nuclear Physics, Moscow State University, Moscow 119992, Russia}
\begin{abstract}
Collisions of metastable antiprotonic helium with medium atoms induce
transitions between hyperfine structure sublevels as well as shift and
broadening of the microwave M1 spectral lines. We consider these
effects in the framework of a simple model with scalar and tensor
interactions between $(\bar{p}\mathrm{He}^+)_{nL}$ and $\mathrm{He}$
atoms. $S$-matrix is obtained by solving coupled-channels equations
involving 4 HFS sublevels $(F=L\pm 1/2,\,J=F\pm 1/2)$ of the $nL$ level
and relative angular momenta up to $l=5$ at the kinetic energy
$E\lesssim 10$ K. The calculated spin-flip cross sections are less than
elastic ones by 4  orders of value. At the density $N=3\times 10^{20}$
cm$^{-3}$ and $T=6$ K we obtain the relaxation times
$\tau(FJ\rightarrow F'J')\gtrsim 160$ ns, the frequency shifts of M1
spectral lines $\Delta\nu\lesssim 66$ KHz for the favored transitions
($\Delta F=\pm 1,\, \Delta J=\pm 1$) and frequency broadening of the M1
spectral lines $\gamma/2\lesssim 5.8$ MHz. The results are compatible
with the recent experimental data obtained by a laser-microwave-laser
resonance method.
\end{abstract}
\end{frontmatter}

The discovery of antiproton longevity in helium and investigations of
antiprotonic helium $(\bar{p}\mathrm{He}^+)$ metastable states by the
method of laser resonance spectroscopy have opened a new layer of very
interesting physics (see \cite{1} and the references therein). One of
the actual problems in the topics is an influence of ambient atoms on
the antiprotonic states. In particular, the density shifts and
broadenings of E1 spectral lines $(nL\rightarrow n'L'=L\pm 1)$ were
observed for laser-induced transitions. The model theoretical analysis
\cite{2} shows that qualitative peculiarities of the data are related
to quantum effects at very low temperature and peculiar features of
$(\bar{p}\mathrm{He}^+)-\mathrm{He}$ interaction. More sophisticated
calculations with \emph{ab initio} potential surface \cite{3} give a
quantitative agreement with the experimental data for a lot of E1
transitions.

Recently, the first data on hyperfine structure of the $(n,L)=(37,35)$
state of $(\bar{p}\mathrm{He}^+)$ were obtained by a
laser-microwave-laser resonance method \cite{4}. The central
frequencies of microwave M1-transitions, $\nu_{HF}^+(F=L-1/2,J=L
\rightarrow L+1/2,L+1)=12.89596(34)$ GHz and $\nu_{HF}^-(L-1/2,L-1
\rightarrow L+1/2,L)=12.92467(29)$ GHz, are in excellent agreement
($\lesssim 30$ ppm) with the recent calculations for the isolated
$(\bar{p}He^+)$ system. The results suppose that density shifts of the
M1 spectral lines at the experimental conditions are very small and do
not exceed the experimental accuracy ($\sim 30$ kHz), in contrast with
the $E1$ transitions, but the width of the lines ($\gamma\sim 5.3\pm
0.7$ MHz) leaves room for a collisional broadening. One more
consequence of the data is that the mean time of collisional relaxation
$\tau_c(F=L-1/2\rightarrow F'=L+1/2)\gtrsim 160$ ns, the latter number
being the observation time window. As far as we know there is no
published theoretical papers on collisional effects on HFS states of
the $(\bar{p}\mathrm{He}^+)$ system. We consider the problem in the
frame of a simple model of $(\bar{p}\mathrm{He}^+)$ - $\mathrm{He}$
potential \cite{2} extended to the tensor interaction.

Hyperfine splitting of the $nL$ levels arises from interactions between
orbital and spin magnetic momenta \cite{5}. A coupling of orbital
angular momentum $\mathbf{L}$ with spin of electron $\mathbf{s}_e$
splits each $nL$ level into two sublevels $F=L\pm 1/2$ with a
displacement $\nu(F_-,F_+)\sim 13$ GHz, each one, due to a coupling
with spin of antiproton $\mathbf{s}_{\bar{p}}$ being in turn split into
two subsublevel $J=F\pm 1/2$ with $\nu(FJ_+,FJ_-)\sim 150$ MHz. To a
first approximation, total wave functions of the split states is
constructed by means of vector coupling of the angular momenta,
$|nLFJM\rangle =\left( \left(\Psi_{nL}(\xi)\otimes\chi_e\right)_F
\otimes \chi_{\bar{p}}\right)_{JM}$, where $\Psi_{nL\Lambda}(\xi)$ is a
spatial wave function, $\chi_e$ and $\chi_{\bar{p}}$ are spin wave
functions.

Interaction between antiprotonic helium in the state
$|nL\Lambda\rangle$ and He atom with account for a tensor term is a
matrix with respect to quantum numbers $\Lambda$ of projection of
angular momentum $L$,
 \begin{equation} \label{1}
 \langle nL\Lambda|V(\mathbf{R},\xi)|nL\Lambda'\rangle=
 V_0(R) \delta_{\Lambda\Lambda'} + V_2(R) \sum_{\nu}
 \langle L\Lambda' 2\nu|L\Lambda\rangle
 Y^*_{2\nu}(\Omega)\sqrt{4\pi/5},
  \end{equation}
where $\mathbf{R}=(R,\Omega)$ is a vector between centers of mass of
the subsystems. Functions $V_0(R)$ and $V_2(R)$ depend on the quantum
numbers $n,L$. For the scalar term we use \cite{2} $V_0(R)= -C_6 f(R)$
with $f(R)= (R^2-r_c^2)/(R^2+r_0^2)^4$ that has a repulsion at $R <
r_c$, Van der Waals minimum at $R^2_{\min}=(r_0^2+4r_c^2)/3$ and  the
correct asymptotic $V_0(R) \rightarrow -C_6/R^6$ at
$R\rightarrow\infty$. Radial dependence of the tensor term at large
distance is similar, $V_2(R)\rightarrow -G_6/R^6$, whereas at small
distance it has to be $V_2(R)\sim R^2$. To satisfy these limits we
suppose $V_2(R) = -G_6 f(R)\cdot[1-\exp(-\eta R^2]$ with a large
$\eta\sim 10$ a.u. For the  calculations we use two sets of the
parameters. The set A is based on the fitting of E1 shifts \cite{2},
$C_6=2.82$, $r_c= 3.0$, $r_0=2.8$ (all values in atomic units). The
second set (B) is estimated using data of \emph{ab initio} calculations
of the potential  energy surface \cite{3,6}. A repulsion radius,
position and depth of Van der Waals minimum of the potential $V_0(R)$
from Fig. 3 in \cite{3} were used to obtain the values $C_6=3.35$,
$r_c= 4.75$, $r_0=0.707$ (a.u.) for our form of the potential. For the
both sets we adopt $G_6/C_6=-0.37$ estimated by means of a
single-particle model of $(\bar{p}\mathrm{He}^+)$ with effective
charges. A dependence of the parameters on $n,L$ is rather weak and
does not matter for our aims. On other hand, two sets of the parameters
differ markedly, that allows to reveal general properties of
characteristics to be considered.

\begin{figure}[hb]
      \centering
\includegraphics[width=0.75\textwidth]{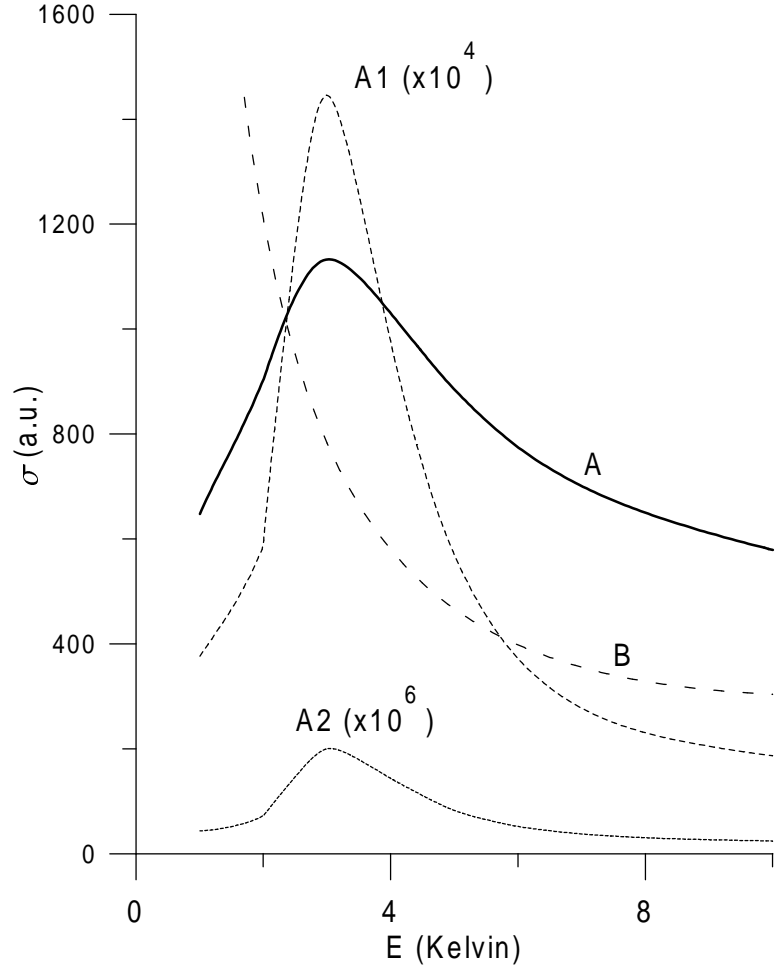}
\caption{Cross sections of $(\bar{p}\mathrm{He}^+)_{nL}$ -
$\mathrm{He}$ collisions. Curves \emph{A} and \emph{B} are total cross
sections averaged over HFS quantum numbers for the sets A and B of the
parameters. Curves \emph{A1} and \emph{A2} are cross sections of
transitions from $F,J=L-1/2,L$ to $L+1/2,L+1$ and $L+1/2,L$ multiplied
by the factors $10^4$ and $10^6$, respectively.} \label{Fig1}
\end{figure}

\begin{figure}[th]
      \centering
\includegraphics[width=0.75\textwidth]{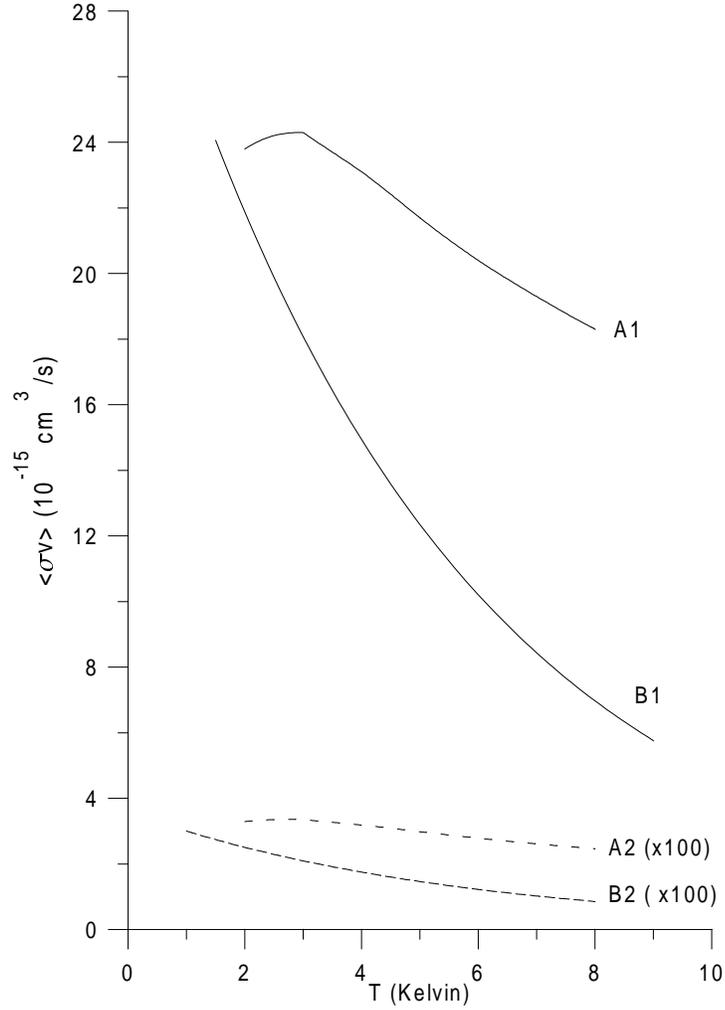}
\caption{Per-atom rate $\langle \sigma v\rangle$ of collisional
transitions averaged over thermal motion depending on temperature.
Curves A1 and A2 are for transition from  $F=L-1/2,J=L$ to $L+1/2,L+1$
and $L+1/2,L$ with the parameters set A, curves B1 and B2 are for the
same transitions with the set B, respectively.} \label{Fig2}
\end{figure}

Second term in \eqref{1} is non-diagonal also with respect to HFS
quantum numbers $FJ$ and to a relative angular momentum $l$ of the
interacting subsystems. Let $|nLFJ,l:jm\rangle = \left(|nLFJ\rangle
\otimes Y_{l}(\Omega)\right)_{jm}$ be a state  with total angular
momentum $\mathbf{j}=\mathbf{J} +\mathbf{l}$ of the whole system. A
matrix of the potentials \eqref{1} on this basis is reduced to the $3j$
and $6j$-symbols,
 \begin{equation} \label{2}
  \begin{split}
 \langle nLFJ,l:jm|V(\mathbf{R},\xi)|nL F'J',l':j'm'\rangle =
  V_0(R) \cdot \delta_{jj'}\delta_{mm'} \delta_{FF'}\delta_{JJ'}\delta_{ll'} \\
 +\quad  V_2(R) \cdot \delta_{jj'}\delta_{mm'} \hat{L} \hat{F} \hat{F'} \hat{J}
\hat{J'} \hat{l} \hat{l'} (-1)^{j+F+F'+L+1} \\
 \cdot \left(\begin{matrix} l'& 2 & l \\ 0 & 0 & 0 \end{matrix} \right)
   \left\{\begin{matrix}J & l & j \\ l' & J' & 2 \end{matrix}\right\}
\left\{\begin{matrix}F & J & 1/2 \\ J' & F' & 2 \end{matrix}\right\}
\left\{\begin{matrix}L & F & 1/2 \\ F' & L' & 2 \end{matrix}\right\} ,
   \end{split}
 \end{equation}
where $\hat{a}\equiv \sqrt{2a+1}$. With these potentials, we solve
coupled channels equations including 4 HFS sublevels at fixed $nL$ and
relative angular momenta up to $l=5$ at the kinetic energy $E\lesssim
10$ K. Elastic and inelastic cross sections and rates of collisional
transitions between HFS states are
\begin{align}
\sigma(FJ\rightarrow F'J')&=\frac{\pi}{k^2} \sum_{jll'}
\frac{2j+1}{2J+1}\left|\delta_{ll'}\delta_{JJ'}\delta_{FF'} -
 \langle FJl|S^j|F'J'l'\rangle \right|^2, \label{3}       \\
\lambda(FJ\rightarrow F'J') &= N \langle v \sigma(FJ\rightarrow
F'J')\rangle,           \label{4}
\end{align}
where N is atomic density of the medium, and angular brackets stand for
averaging over thermal velocity distribution. The results for the cross
sections are shown on the Fig. \ref{Fig1}. Non-diagonal elements of the
S-matrix generated by the tensor term in \eqref{1} are small, therefore
spin-flip cross sections are less than elastic one by 4 orders of
value. Fig. \ref{Fig2} shows the calculated per-atom collisional
transition rates averaged over the thermal velocity distribution.

\begin{figure}[th]
      \centering
\includegraphics[width=0.75\textwidth]{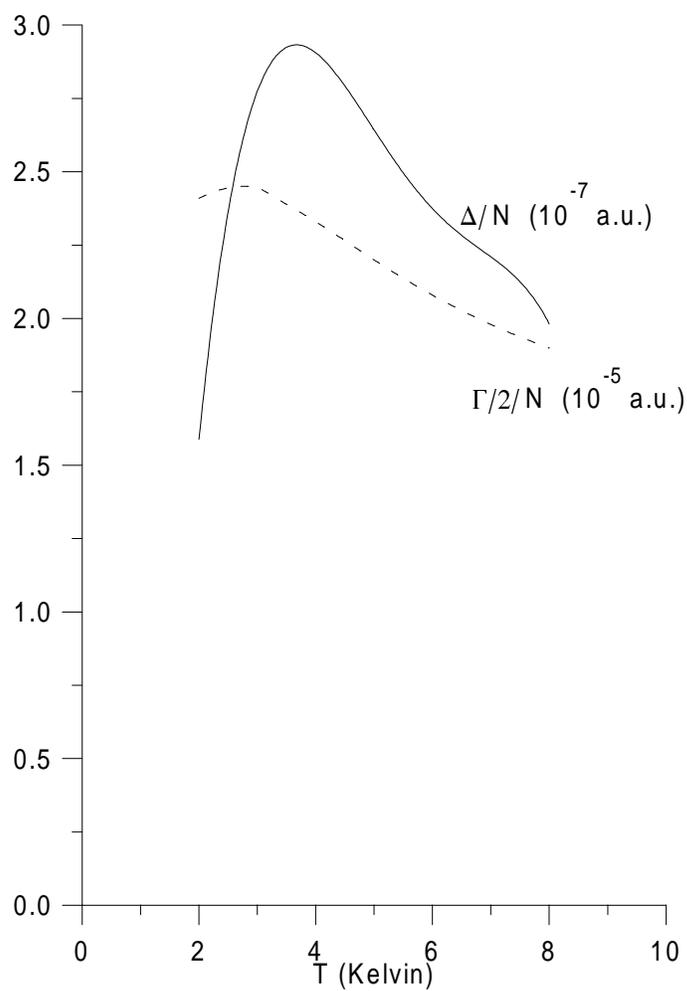}
\caption{Per-atom collisional shift and broadening of the M1 spectral
line  $\Delta(F=L-1/2, J=L \rightarrow L+1/2, L+1)$ depending on
temperature.}  \label{Fig3}
\end{figure}

Collisional shifts and broadenings of the microwave M1 transitions can
be considered using a general theory of similar effects in atoms. For
non-overlapping levels, Eq. (57.96) from \cite{7} is relevant to our
problem, and, with our notations of the quantum numbers, becomes
\begin{multline} \label{5}
\gamma + \mathrm{i}\Delta = N\pi \sum_{ll'j_1j_2}
(2j_1+1)(2j_2+1)(-1)^{l+l'}
 \left\{\begin{matrix} j_1 & j_2 & 1 \\ J_2 & J_1 & l \end{matrix}\right\}
 \left\{\begin{matrix} j_1 & j_2 & 1 \\ J_2 & J_1 & l'\end{matrix}\right\}
  \\
 \cdot  \langle vk^{-2} [\delta_{ll'} -  \langle nLF_1J_1 l'|S^{j_1}|nLF_1J_1 l\rangle
  \langle nLF_2J_2 l'|S^{j_2}|nLF_2J_2 l\rangle ^*]  \rangle,
\end{multline}
where S-matrix with the quantum numbers subscript 1 (2) corresponds to
collisions before (after) M1 transition $(F_1 J_1\rightarrow F_2 J_2)$.
Fig. \ref{Fig3} shows temperature dependence of shift and broadening
for the transition $F=L-1/2, J=L\rightarrow L+1/2,L+1$ calculated with
the set A of the parameters.

At the experimental conditions ($T=6$ K, $N=3\cdot10^{20}$ cm$^{-3}$)
we obtain from Fig. \ref{Fig2} a relaxation rate
 $\lambda(F_- \rightarrow F_+)\lesssim 6\cdot 10^6$ s, or $\tau\gtrsim 160$ ns. At
the same conditions, shift and broadening of the M1 line from Fig.
\ref{Fig3} are estimated as $\Delta\nu=65.8$ KHz and $\gamma/2=5.8$
MHz. In spite of rather simple used model, the results can be
considered as compatible with upper limits of these value from the
recent experimental data.

   \end{document}